
\input vanilla.sty 
\font\tenbf=cmbx10
\font\tenrm=cmr10
\font\tenit=cmti10
\font\ninebf=cmbx9
\font\ninerm=cmr9
\font\nineit=cmti9

\font\eightrm=cmr8
\font\eightit=cmti8
\font\sevenrm=cmr7
\TagsOnRight
\hsize=5.0truein
\vsize=7.8truein
\parindent=15pt
\nopagenumbers
\baselineskip=10pt
\def\qed{\hbox{${\vcenter{\vbox{
    \hrule height 0.4pt\hbox{\vrule width 0.4pt height 6pt
    \kern5pt\vrule width 0.4pt}\hrule height 0.4pt}}}$}}
\line{\eightrm International Journal of Modern Physics C, to be published
\hfil}
\line{\eightrm presented at the workshop "Dynamics of First Order Phase
Transitions, June 1-3, J\"ulich, Germany \hfil}
\vglue 5pc
\baselineskip=13pt
\centerline{\tenbf ON THE TWO-POINT CORRELATION FUNCTION IN}
\centerline{\tenbf DYNAMICAL SCALING AND SCHR\"ODINGER INVARIANCE}
\vglue 24pt
\centerline{\eightrm MALTE HENKEL}
\baselineskip=12pt
\centerline{\eightit Department of Theoretical Physics, University of Geneva}
\baselineskip=10pt
\centerline{\eightit 24 quai Ernest Ansermet, CH-1211 Geneva 4, Switzerland}
\vglue 20pt
\baselineskip=10pt
\vglue 16pt
\centerline{\eightrm ABSTRACT}
{\rightskip=1.5pc
 \leftskip=1.5pc
 \eightrm\baselineskip=10pt\parindent=1pc
The extension of dynamical scaling to local, space-time
dependent rescaling factors is investigated. For a dynamical
exponent $z=2$, the corresponding invariance group is the
Schr\"odinger group. Schr\"odinger invariance is shown to
determine completely the two-point correlation function. The
result is checked in two exactly solvable models.
\vglue 5pt
\noindent
{\eightit Keywords}\/: Dynamical scaling; Glauber dynamics;
Schr\"odinger invariance; conformal invariance
\vglue 12pt}
\baselineskip=13pt
\line{\tenbf 1. Introduction \hfil}
\vglue 5pt
The concept of scaling has proved to be a very fruitful one in describing
phase transitions of statistical systems. For {\it static} critical
phenomena, the renormalization group (see e.g. Ref.~1) has elucidated the
universal quantities characterizing a universality class and has provided
approximation schemes for their calculation. More recently, conformal
invariance, at least in two dimensions, has yielded exact results for
critical exponents and amplitude ratios and also for the multipoint
correlation functions (see e.g. Ref. 2).

Much less is known for time-dependent problems$^{3}$.
However, it was recognised that
dynamical scaling may arise in various situations, that is, the two-point
correlation function $C(\vec{r},t)$ satisfies
$$ C( \lambda \vec{r}, \lambda^{z} t) = \lambda^{-2x} C(\vec{r},t) \tag 1.1$$
where $z$ is the dynamic critical exponent and $x$ is the scaling dimension.
Examples include the time-dependent behaviour at a static critical
point$^{1,3}$
or the ordering process following the quench of a system from an initial state
at high temperatures to a final state below the critical temperature$^{4}$.
In this case, the corresponding static system is not critical. For a recent
experimental example (with conserved order parameter) for the development of
dynamical scaling at late times in
${\tt Mn}_{0.67}{\tt Cu}_{0.33}$, see Ref.~5.

Eq.~(1.1) can be recast into the form
$$ C(r,t) = t^{-2x/z} \Phi\left( \frac{r^{z}}{t} \right) \tag 1.2$$
defining the scaling function $\Phi(u)$ where $u=r^{z}/t$ is the scaling
variable. What can be said about $\Phi(u)$ ?

We propose to generalize the dynamic scaling (1.1) from the global form with
$\lambda = \text{ const.}$ to a {\it local} one, where the rescaling
factor becomes space-time dependent
$$ \lambda = \lambda (\vec{r},t) .\tag 1.3$$
This is analogous to the introduction of conformal
invariance in statics$^{6}$. In
fact, the direct generalization of conformal invariance to dynamics, with
$\lambda=\lambda(\vec{r})$, was attempted$^{7}$ some time ago, being restricted
to two space dimensions and to the case where the static system
is at a second order critical point. The approach to be presented here is not
subject to any of these restrictions.

Here, we shall concentrate on the special case $z=2$ with a non-conserved order
parameter. The group of local scale transformations is then the
Schr\"odinger group, to be defined in the next section. We shall show that
$$ \Phi(u) = \Phi_{0} \exp \left( - \frac{1}{2} {\cal M} u \right) \tag 1.4$$
where $\cal M$ and $\Phi_0$ are constants. This result will be confirmed
in section~3 for the $d$-dimensional spherical model with a non-conserved
order parameter$^{8}$ and for the one-dimensional Glauber-Ising model$^{9}$.
In section~4, we conclude with a brief outlook for the case $z\neq 2$.
\vglue 12pt
\line{\tenbf 2. The Schr\"odinger group \hfil}
\vglue 5pt
The Schr\"odinger group $Sch(d)$ in $d$ space dimensions is defined$^{10,11}$
by
the following space-time transformations
$$ \align \vec{r} \rightarrow \vec{r'} &= \frac{{\cal R} \vec{r} + \vec{v}\, t
+
\vec{a}}{\gamma t + \delta} \tag 2.1 \\
t \rightarrow t' &= \frac{\alpha t + \beta}{\gamma t + \delta}
\;\; ~~~~~~~~~~ \; ; \; ~~~ \;\;
\alpha \delta - \beta \gamma =1 \endalign$$
where $\cal R$ is a rotation matrix and $\alpha, \beta, \gamma, \delta,
\vec{v}, \vec{a}$ are parameters. The Schr\"odinger group is the maximal group
which transforms$^{10}$ solutions of the Schr\"odinger equation
$$ \left( i \frac{\partial}{\partial t} + \frac{1}{2m} \sum_{j=1}^{d}
\frac{\partial^{2}}{\partial r_j \partial r_j} \right) \psi =0 \tag 2.2$$
into other solutions of (2.2), viz. $(\vec{r},t) \mapsto g(\vec{r},t)$,
$\psi \rightarrow T_g \psi$
$$ \left( T_g \psi \right) (\vec{r},t) = f_{g}\left( g^{-1}(\vec{r},t) \right)
\psi\left( g^{-1}(\vec{r},t) \right) \tag 2.3$$
where $f_g$ is the companion function which has been worked out explicitly. For
diffusive processes, (2.2) may be replaced by the Helmholtz equation, with
$2im = D^{-1}$, where $D$ is the diffusion constant. It is also known that
Euclidean free field theory is invariant$^{11}$ under
(2.3). In particular, choosing
$\vec{v}=\vec{a}=0$, $\beta=\gamma=0$ and $\alpha=1/\delta$, we recover the
global scale transformation $\vec{r} \rightarrow \alpha \vec{r}$, $t
\rightarrow
\alpha^{2} t$, which corresponds to $z=2$. For simplicity, we take $d=1$ in the
sequel, but all the results to be described here generalize immediately to
arbitrary $d$.

The infinitesimal generators are
$$ \align X_n &= - t^{n+1} \, \partial_t - \frac{n+1}{2} t^n r \partial_r -
\frac{n(n+1)}{4} {\cal M} t^{n-1} \, r^{2} \;\; ; \;\; n = -1,0,1 \tag 2.4a\\
Y_n &= - t^{n+1/2} \, \partial_r - \left( n+\frac{1}{2} \right) {\cal M}
t^{n-1/2} \, r
\; ~~~~~~ \;  ; \;\; n = -\frac{1}{2}, \frac{1}{2} \tag 2.4b\\
M_n &= - t^{n} {\cal M} \;\;  ; \;\; n=0 \tag 2.4c\endalign$$
where the terms $\sim {\cal M}$ in $X_n$ and $Y_n$ come from the companion
function. The commutation relations are
$$ \align
\left[ X_n , X_m \right] &= (n-m) X_{n+m} \\
\left[ X_n , Y_m \right] &= \left( \frac{n}{2} - m \right) Y_{n+m} \\
\left[ X_n , M_m \right] &= - m M_{n+m} \tag 2.5\\
\left[ Y_n , Y_m \right] &= (n-m) M_{n+m} \\
\left[ Y_n , M_m \right] &= \left[ M_n , M_m \right] = 0 .\endalign$$
and it follows that the set
$\{ X_{-1}, X_{0}, X_{1}, Y_{-1/2}, Y_{1/2}, M_{0}\}$
spans a six-dimensional subalgebra.
In order to implement Schr\"odinger invariance on the
correlation functions, we have to
replace in (1.1) the factor $\lambda^{-2x}$ by the corresponding Jacobian
$$ < \phi_{1}({r'}_{1},t_{1}') \ldots \phi_{n}({r'}_{n}, t_{n}')> =
\prod_{i=1}^{n} \left|
\frac{\partial({r'}_{i},t_{i}')}{\partial({r}_{i},t_{i})}\right|^{-x_{i}/(2+d)}
< \phi_{1}({r}_{1},t_{1}) \ldots \phi_{n}({r}_{n}, t_{n})>  .\tag 2.6$$
This is the analogue of the definition of the quasiprimary fields of conformal
invariance. In particular, derivative fields are excluded by (2.6).

We now examine the consequences for the two-point function
$$ F(r_1 , r_2 ; t_1 , t_2 ) =
< \phi_1 (r_1, t_1), \phi_2 (r_2 , t_2 ) > .\tag 2.7$$
Translations are generated by $X_{-1}$ and $Y_{-1/2}$ and imply that
$F=F(r_1 - r_2 , t_1 - t_2)$. Global scale transformations
are generated by $X_0$.
Writing $r=r_1 - r_2$ and $t=t_1 - t_2$, we have
$$ \left( t \partial_t + \frac{1}{2} r \partial_r + \frac{1}{2}
\left( x_1 + x_2 \right) \right) F = 0 \tag 2.8$$
yielding, with $x=x_1 + x_2$
$$ F(r,t) = t^{-x/2} G(u) \;\; ; \;\; u = r^2 /t \tag 2.9$$
and reproducing (1.2). The new information comes from the Galilei
transformation, generated by $Y_{1/2}$. By translation invariance, we can put
$r_2 = t_2 =0$. Then
$$ \left( t \partial_r + {\cal M} r \right) t^{-x/2} G(u) =0 \tag 2.10$$
which gives an equation for $G(u)$
$$ \frac{d G}{d u} + \frac{\cal M}{2} G = 0 \tag 2.11$$
and thus, with $\Phi_0$ being a normalization constant
$$ F(r,t) = \Phi_0 t^{-x/2} \exp\left( -
\frac{\cal M}{2} \frac{r^2}{t} \right) .\tag 2.12$$
We still have to see whether this is consistent with the special Schr\"odinger
transformation generated by $X_1$. Using translational invariance, we put
$r_2 = t_2 =0$ and have
$$ \left( t^2 \partial_t + t r \partial_r
+ \frac{\cal M}{2} r^2 + 2 t \frac{x_1}{2}
\right) t^{-x/2} G(u) = 0 \tag 2.13$$
and we can see that this implies two conditions.
The first one is just (2.11), while
the second one is $x=2x_1$ or, since $x=x_1 + x_2$
$$ x_1 = x_2 \tag 2.14$$
which means that the two scaling operators $\phi_1 , \phi_2$ have to be the
same in order to have a non-vanishing two-point correlation function.
We summarize our result
$$ < \phi_{1}(\vec{r},t) \phi_{2}(\vec{0},0)> = \delta_{1,2} \Phi_0 t^{-x_1}
\exp \left( - \frac{\cal M}{2} \frac{|\vec{r}\, |^{2}}{t} \right) \tag 2.15$$
where we have restored the $d$-dimensional case.
For the special case $d=2$, this
is in agreement with the conformal invariance approach of Ref. 7.
Comparing with the corresponding result of conformal invariance$^{6}$ of
a static critical point, we note
the importance of the contribution arising from the companion term in the
Schr\"odinger group which is parametrized by the non-universal
constant $\cal M$.

One may go on and consider higher correlation functions. Furthermore,
one can show
that invariance under translations, dilatations, space rotations and Galilei
transformations imply the full Schr\"odinger invariance if the interactions
are short-ranged. This will be presented in detail elsewhere$^{12}$.
\vglue 12pt
\line{\tenbf 3. Comparison with exactly solvable models \hfil}
\vglue 5pt
We now compare the result for the two-point function (1.4,2.15) with two
exactly solvable time-dependent models which have $z=2$.

The first model we consider is the $O(N)$-symmetric
time-dependent Ginzburg-Landau model$^8$.
Initially, the system is at very high temperatures, but at time $t=0$, it
is quenched to zero temperature. In the $N\rightarrow \infty$ limit, the
structure function was calculated exactly$^{8}$ for late times in $d$
spatial dimensions
$$ C(\vec{k},t) = M_{0}^{2} L^{d}(t)
\exp\left( - k^2 L^{2}(t) \right) \tag 3.1$$
where $M_0$ is the equilibrium magnetization,
$L(t) = (2Dt)^{1/2}$ is the typical
domain size and $D$ is the diffusion constant. This can be rewritten in
direct space
$$ C(\vec{r},t) = 2^{-d/2} M_{0}^{2} \exp\left( - \frac{d}{8D}
\frac{|\vec{r}\, |^{2}}{t} \right)
\tag 3.2$$
in agreement with (1.4,2.15) and we read off
$x=0$. Since the renormalization group
eigenvalue $y=d-x$ and $y=d$ at a first-order transition, this last result is
probably not too surprising.

As a second example, we take the one-dimensional Ising model with Glauber
dynamics$^{9}$. If the system is in thermal equilibrium at temperature $T$, the
spin-spin correlation function is known exactly$^{9}$ ($t > 0$)
$$ \align C(a-b,t) &= < \sigma_{a}(0) \sigma_{b}(t)> \\
 &= e^{-\alpha t} \sum_{\ell =
-\infty}^{\infty} \eta^{|a-b+\ell|} I_{\ell}(\gamma
\alpha t) \tag 3.3\endalign$$
where $\alpha$ is the transition rate, $\eta=\tanh J/k_B T$,
$\gamma = \tanh 2J /k_B T$, $J$ is the exchange integral of the Ising model
and $I_{\ell}$ is a modified Bessel function. To analyse this, we recall
the asymptotic expansion$^{13}$, as $x\rightarrow \infty$
$$I_{\ell}(x) \simeq (2\pi x)^{-1/2} \exp\left( x - \frac{\ell^2}{2x} \right)
\left( 1 + O(x^{-1}) \right) \tag 3.4$$
and, writing $r=a-b$, we have
$$C(r,t) \simeq e^{-\alpha(1-\gamma)t} (2\pi \gamma \alpha t)^{-1/2}
\left\{ \exp\left( -\frac{r^2}{2\gamma\alpha t}\right) +
\sum_{\ell \neq 0} \eta^{|\ell |}
\exp\left( - \frac{(r+\ell)^{2}}{2\gamma\alpha t}
\right) \right\} .\tag 3.5$$
Now take the simultaneous scaling limit
$r\rightarrow\infty$, $t\rightarrow\infty$
such that $u = r^2 /t$ stays fixed. Then the leading term becomes
$$ C(r,t) \sim e^{-\alpha(1-\gamma)t} (2\pi\gamma\alpha t)^{-1/2}
\exp \left( - \frac{1}{2\gamma\alpha} \frac{r^2}{t} \right) \tag 3.6$$
where each of the terms neglected is an exponentially small
correction-to-scaling term.

The first factor in (3.6) describes the off-critical relaxation towards
equilibrium and we can identify the well-known relaxation time
$\tau^{-1} = \alpha (1-\gamma)$. At the critical point $T=0$, we have
$\gamma=1$
and this factor becomes unity. Alternatively, we can
define another temperature-time
scaling limit, where $t\rightarrow\infty$, $T\rightarrow 0$ such that
$t \exp - 4J/k_{B} T$ is kept fixed. We then
recover the anticipated scaling form
and find agreement with (1.4,2.15). We read off $2x=1/2$.
\vglue 12pt
\line{\tenbf 4. Conclusions and outlook \hfil}
\vglue 5pt
The hypothesis of Schr\"odinger invariance in critical dynamics with $z=2$
was shown to predict the scaling function of
the two-point correlation function in the case of a non-conserved order
parameter. This finding is supported by results from exactly solvable models.
The theory will be developed more systematically elsewhere$^{12}$.

We comment briefly on the possibility to generalize
beyond the case $z=2$. It can be argued$^{12}$ that for $u$ large
$$ \Phi(u) \sim \exp \left( - \text{ const.} \, u^{1/(z-1)} \right) .\tag 4.1$$
We are not aware of any calculation in critical dynamics which either supports
or excludes (4.1). However, (4.1) is supported in static, but strongly
anisotropic systems, where now $\theta=\nu_{\|}/\nu_{\perp}$ measures the
anisotropy and enters in the scaling form (1.1), (1.2) instead of $z$.
Examples are provided by Lifshitz points in the spherical model$^{14}$
($\theta=1/2, 2, 3$) and two-dimensional directed percolation$^{15}$
($\theta \simeq 1.58$).
Conformal invariance, valid only for $d=2$ and at an isotropic (or static)
critical point, suggests
that$^{7}$ $\Phi(u) \sim e^u$ independent
of $z$ or $\theta$. Altough this result is in
agreement with ours for $\theta=2$ (or $z=2$), it is at variance with (4.1)
if $\theta \neq 2$. Unfortunately, in none of
the models with $\theta \neq 2$ studied
so far the correlation function was calculated in $d=2$ so that a direct
comparison has not yet been possible. A detailed account will be given in
a separate paper$^{12}$.
\vglue 12pt
\line{\tenbf Acknowledgements \hfil}
\vglue 5pt
I thank the organizers for the stimulating environment during the workshop and
M. Droz, L. Frachebourg and A. Patk\'os for useful discussions.
Early parts of this
work were done at the Institute of Atomic Physics at E\"otv\"os University,
the visit being made possible by a grant of the Swiss National Science
Foundation in the framework of the Program of Cooperation with Eastern
European Countries. This work was further supported
by the Swiss National Science Foundation.
\vglue 12pt
\line{\tenbf References \hfil}
\vglue 5pt

\medskip
\ninerm
\baselineskip=11pt
\frenchspacing
\item{1.} S. K. Ma {\nineit Modern Theory of Critical Phenomena}
(Benjamin, 1976).
\item{2.} J. L. Cardy in E. Br\'ezin and J. Zinn-Justin (Eds.) {\nineit
Fields, strings and critical phenomena}, Les Houches XLIX, (North Holland,
1990), p. 169
\item{3.} B.I. Halperin and P. C. Hohenberg {\nineit Rev. Mod. Phys.}
          {\ninebf 49}, 435 (1977)
\item{4.} K. Binder and D. Stauffer {\nineit Phys. Rev. Lett.} {\ninebf 33},
          1006 (1974);
J. Marro, J. Lebowitz and M. H. Kalos {\nineit Phys. Rev. Lett.} {\ninebf 43},
          282 (1979)
\item{5.} B. D. Gaulin, S. Spooner and Y. Morii {\nineit Phys. Rev. Lett.}
          {\ninebf 59}, 668 (1987)
\item{6.} A. M. Polyakov {\nineit Sov. Phys. JETP Lett.} {\ninebf 12},
          381 (1970)
\item{7.} J. L. Cardy {\nineit J. Phys.} {\ninebf A18}, 2771 (1985)
\item{8.} A. Coniglio and M. Zanetti {\nineit Europhys. Lett.}
          {\ninebf 10}, 575 (1989)
\item{9.} R. Glauber {\nineit J. Math. Phys.} {\ninebf 4}, 294 (1963)
\item{10.} U. Niederer {\nineit Helv. Phys. Acta} {\ninebf 45}, 802 (1972)
\item{11.} C. R. Hagen {\nineit Phys. Rev.} {\ninebf D5}, 377 (1972)
\item{12.} M. Henkel {\nineit to be published}
\item{13.} S. Singh and R. K. Pathria {\nineit Phys. Rev. } {\ninebf B31},
           4483 (1985)
\item{14.} L. Frachebourg and M. Henkel {\nineit to be published}
\item{15.} J. Benzoni {\nineit J. Phys.} {\ninebf A17}, 2651 (1985)

\vfil\supereject
\bye